\documentclass[sigconf,screen]{acmart}
\setcopyright{acmcopyright}
\acmPrice{15.00}
\acmDOI{10.1145/3338906.3341458}
\acmYear{2019}
\copyrightyear{2019}
\acmISBN{978-1-4503-5572-8/19/08}
\acmConference[ESEC/FSE '19]{Proceedings of the 27th ACM Joint European Software Engineering Conference and Symposium on the Foundations of Software Engineering}{August 26--30, 2019}{Tallinn, Estonia}
\acmBooktitle{Proceedings of the 27th ACM Joint European Software Engineering Conference and Symposium on the Foundations of Software Engineering (ESEC/FSE '19), August 26--30, 2019, Tallinn, Estonia}

\usepackage[utf8]{inputenc}
\usepackage{subcaption}

\begin{document}

\title{Exploratory Test Agents for Stateful Software Systems}

\author{Stefan Karlsson}
\email{stefan.l.karlsson@{se.abb.com, mdh.se}}
\affiliation{
  \institution{ABB AB, Mälardalen University}
  \city{Västerås}
  \country{Sweden}
}

\begin{abstract}
The adequate testing of stateful software systems is a hard and costly activity.
Failures that result from complex stateful interactions can be of high impact, and it can be hard to replicate failures resulting from erroneous stateful interactions.
Addressing this problem in an automatic way would save cost and time and increase the quality of software systems in the industry. In this paper, we propose an approach that uses agents to explore software systems with the intention to find faults and gain knowledge. 
\end{abstract}



\maketitle
\section{Introduction}
In my experience, from working over a decade in the software industry there is a lack of industry-strength methods to automatically test and explore the properties of stateful systems and processes. Further, to the extent that such methods do exist in academia, they have not reached wide acceptance in the industry.

Developing and maintaining automatic test cases is a lot of work for developers and even if this is done, the input domain of any non-trivial system is not even close to being covered. Consequently, software projects do not spend the effort needed to create and maintain sufficient automatic tests. In addition, bugs that need more complex interactions than a simple unit-test can find are found at a late stage of testing by a human, or in the worst case, in the field by a customer.

New deployment models for software such as cloud based micro-service architectures make it insufficient to only execute tests on a per artifact basis. Tests must continuously run in an environment where all these interacting services are running to be able to explore complex interactions. 

It follows from that reasoning that tests that aim to expose faults that require stateful software processes to run together for an unknown number of interactions must have more intelligence and autonomy than a static example based approach.

In the same way as garbage collection has reduced the manual developer labor of memory management, there would be developer effectiveness gains in a more automatic approach to testing of software systems. Others have also reported from the field that developers and testers need more ways to automatically generate test cases \cite{Arcuri2018}. Automatically generated test cases would also be a solution for adding tests where there are too few existing tests. The lack of tests could be a result of time constraints or neglect.

This work aims to research the possibility of designing and using exploratory software test agents to reduce the burden of manual testing, increase the coverage of stateful interactions and to thereby increase the knowledge of how the System Under Test (SUT) behaves.

A goal of this research is to be useful for engineers working in the industry today. Therefore, we will try to find, evaluate and introduce existing methods where applicable, integrating them into the proposed autonomous test agents. Examples of such available methods that could be leveraged by agents are model-based testing \cite{doi:10.1002/stvr.456} and property-based testing \cite{Papadakis:2011:PIT:2034654.2034663, Claessen:2011:QLT:1988042.1988046}. At the same time, we do also need to identify where there are gaps in methods and more research is needed.

To be able to explore the SUT an agent needs some way to interact with the SUT. Possible ways are GUI or API:s.
Whichever method is selected, an agent that acts on the exploratory level of testing, i.e. the end user level, will need a model of how a user can behave. It would be desirable to find a model that enables as fast test iterations as possible.
Fast iterations would increase the likelihood that if an agent can find the fault, it will be closer in time to when the fault was introduced lowering the risk of finding errors late in the development cycle where cost is higher to address the error.
In the case of an intelligent agent, that might require training, fast iterations will reduce the total time of the training cycles needed.

Results from running a test agent must be asserted on and available for domain experts to analyze. The results constitute an important source to evaluate if test goals have been reached, both for human observers and test agents. Both agents and humans could use the results for future learning and as a source of knowledge of how the system behaves. Since a goal of this work is to lessen the burden of testing, care must be taken not to move that effort to the formulation of agent goals and assertions or to analyze and reproduce any error state. Thus, any research resulting in common ways of expressing agent interactions and goals and how to reproduce problems is of great value. 

In summary, my research aims at contributing:
\begin{itemize}
    \item Interaction models for test agents.
    \item Recommendations for formulation of assertions of exploratory test results data.
    \item Methods to analyze and learn from exploratory test results.
    \item A method of how to define exploratory agent goals.
\end{itemize}

The importance of those research goals is:
\begin{itemize}
    \item To enable test where other levels of testing is missing.
    \item To find bugs that require more complex interactions to show.
    \item To reduce the human effort of testing and enable humans to focus on more complex interactions.
    \item To empower developer learning of the system under test.
\end{itemize}

\section{Background}
In this work we investigate the feasibility of using exploratory test agents to learn more about the SUT and its environment. Since "exploratory test agents" is not a mainstream term, it is useful to define how these terms are used in this work.

\subsection{Definitions}

\paragraph{Exploratory}
 We do not want to limit ourselves to the domain of exploratory testing and therefore choose a very open definition of exploration. A dictionary definition of "Exploration" is:
"Exploration is the act of searching for the purpose of discovery of information or resources" \cite{WikiExplore}.
A test agent then acting as the "explorer" would help us to discover what we do not know about the SUT. This information includes bugs, what possible interactions exists and the performance properties of the system. It can also include meta-properties of the system. An example of such a property would be if logging is handled in a way to over time not fill up discs.

\paragraph{Test Agents}
What an agent is has been debated and there have been taxonomies proposed of the definition as well as its different classes. Generally, an agent senses and acts according to its goals with different levels of autonomy \cite{10.1007/BFb0013570}. This work focuses specifically on test agents. We share the definition of intelligent test agents as autonomous and more intelligent test cases that can interact with each other and self-select what tests to run \cite{8728945}. 

In the scope of this work we will consider both intelligent agents, in the form of intelligent test cases, and agents lacking intelligence such as agents that sense they should perform a test with a goal but that do not learn or adapt from its environment. Some of the test agents might be considered to be programs since they do not perform changes to its environment, or reach the levels of perception and autonomy expected by agents. However, most of them would have the property of running over a given time. The main goal is to ease the burden of finding faults for software developers and in some cases an agent with no intelligence, that do not learn, will be sufficient to fulfill this objective.

\section{Related Work}

One challenge in this work is how the test agents interact with the SUT. The interaction model can be implicit such as a GUI or an explicit artifact such as a model.

To address the problems of explicit models, such as cost and availability, lots of learning-based testing have been researched in several different domains to learn the models of the SUT. According to Aichernig et al. this introduces the general trend in society to make things smarter to the domain of test \cite{Aichernig2018}.

Marculescu et al. have used several different exploration-based algorithms to augment search-based testing \cite{7515460}. Their results indicate that automatic exploratory testing could be used successfully where there are limitations to the available detailed domain information of the SUT. The usage of automatic exploratory tests could be fully automatic as well as an augmentation to a human expert that directs the exploration.

In recent years there have been tools developed that uses search-based exploration or evolution to fulfill objectives.

In the domain of micro-service based REST-API:s, EvoMaster introduces an approach of automatic test case generation. EvoMaster uses a white-box approach that optimize on the criteria of maximum code coverage and fault-finding \cite{Arcuri:2019:RAA:3292526.3293455}. To learn more about the SUT these approaches could be augmented with more intelligent test cases acting closer to a real user.

For Android applications, model-based approaches have also been augmented with exploration and multiple objectives to successfully find faults. Recent examples of tools where this have been done are AimDroid \cite{8094413}, Sapienz \cite{Mao:2016:SMA:2931037.2931054}, MobiGUITAR \cite{6786194} and Stoat \cite{Su:2017:GSM:3106237.3106298}. Both AimDroid and Sapienz aim to reduce any fault finding interaction sequence. All of the tool's UI interaction is implemented close to the platform and are tied to Android, so more work would be needed to generalize the tools to other platforms. 

To generalize exploratory tests for different kinds of systems and platforms visual GUI testing could be used. According to Alegroth, the applicability of Visual GUI testing has been shown in some initial results but these were done with test scripts and further research is needed for exploratory testing \cite{alegroth15}.

AppFlow is a recent tool that tries to alleviate the burden of UI testing by increasing test re-use and decreasing the brittleness of UI test scripts \cite{Hu:2018:AUM:3236024.3236055}.
AppFlow uses machine learning to recognise screens and to generate generic tests that are reusable from high-level scripts. The tool allows developers to re-use tests written for some category of the app. As an example, the authors suggests "add to cart" as such a category for a range of shopping apps. AppFlow is currently developed for the Android Platform but its techniques can be adopted to other UI platforms. As stated by the authors, AppFlow cannot test Apps that require external interactions such as SMS authentication. For scenarios like this a multi-agent scenario for test agents could be considered. Such that the agent that explores the App receives the SMS code from another agent.

The Augusto tool uses an interesting approach that could be useful for exploration by encoding, what the authors call, "common sense knowledge" as "application independent functionalities", uses the GUI to discover these predefined scenarios and generates tests \cite{8453087}. According to the authors, this helps in reducing the state space explosion of exploration.

Intelligent Test Agents have been considered in the domain of regression testing. In that scenario testing agents would decide what tests that need running given changes to the system \cite{8728945}.

In artificial intelligence research it has been shown that models for learning do not have to be based on extrinsic design rewards, but good agent performance can be achieved with intrinsic rewards such as curiosity \cite{DBLP:journals/corr/abs-1808-04355}. Such approaches could allow intelligent test agents to explore an application without the effort of designing rewards or explicit models. 

\begin{figure}
  \begin{subfigure}{.23\textwidth}
    \centering
    \includegraphics[width=\linewidth]{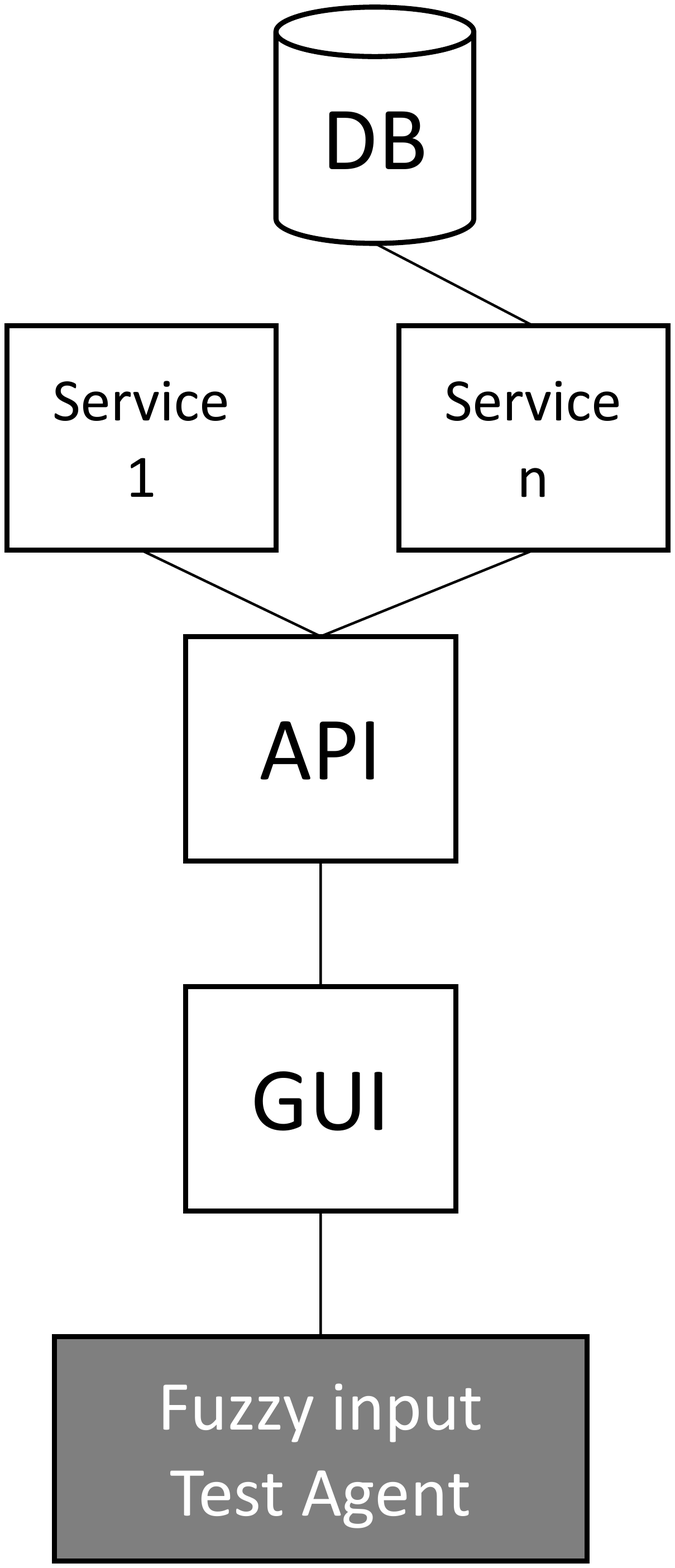}
    \caption{Single agent scenario}
    \label{fig:singleagent}
  \end{subfigure}
  \begin{subfigure}{.23\textwidth}
    \centering
    \includegraphics[width=\linewidth]{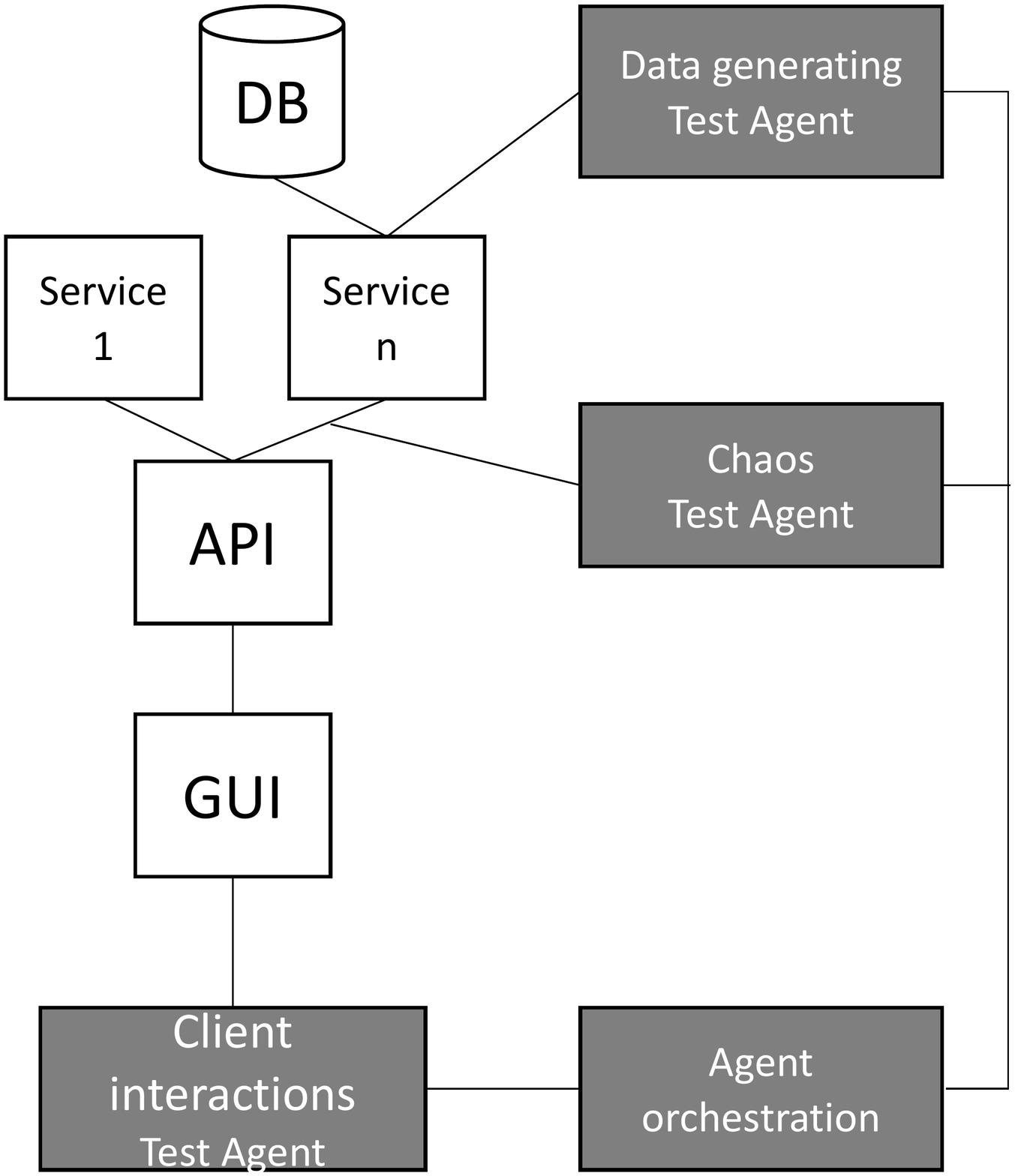}
    \caption{Multiple agent scenario}
    \label{fig:multiagent}
  \end{subfigure}
  \caption{Single- and multi-agent scenarios}
  \label{fig:agents}
  \Description{Single- and multi-agent scenarios}
\end{figure}

\section{Proposed approach}

In this work I envision a group of software agents with varying levels of intelligence that exercise and explore the SUT to expose faults and to increase the knowledge of the system behaviour. This group of agents would help developers shorten the feedback cycle of testing, expose hard to find errors in long running stateful processes and let human testers focus their testing effort on the highest level of exploration and the evaluation of agent output.

This group of agents would then augment the humans in the loop by notifying on unexpected behaviour and ask for human consultation of anomalies.

An example of a multiple agent scenario as described above can bee seen in Figure \ref{fig:multiagent}. For situations where we only want to test one aspect of the system or we might require on offline scenario, we could use a single agent to test that aspect without any connection to other agents, as in Figure \ref{fig:singleagent}.

To make this vision more concrete we can imagine some examples of different kinds of test agents with different kinds of sensors and goals.

\begin{itemize}
    \item The fuzzing agent, who senses a service specification and generates fuzzy input, observing any crashes.
    \item The smoke-test agent, sensing a redeploy of the system, run a basic path exercising the system, observing differences to previous runs.
    \item The chaos agent, stops services and restarts services, introduces latency and observes the resilience of services.
    \item The client agent, explores the system given a client interaction model and tries to reach stated goals.
    \item The security agent, explore services and assess them on security best practices. 
\end{itemize}

These are a few examples of different kinds of test agents but it paints the picture of agents working together to explore the SUT. To be able to work in a group the agents need a common way of expressing results and goals, and most likely test agents that work on a higher level, monitoring the results of lower level agents.

Given this vision the main themes of the approach is then, how can the test agents interact with the SUT? How are goals formulated for a test agent? How do we assert on the result?

\subsection{Agent Interaction}

Any test needs a way to interact with the SUT. For example, a unit test will interact with the SUT by using the public methods or functions defined. Since the test agent should perform exploratory higher-level tests, its interaction model must also be on a higher level.
In MBT this model is separated from implementation. One reasoning behind this is to be able to find faults outside of the current implementation. i.e. requirements that are not implemented correctly or missing.

It is worth considering that implementations and tests always have a shared "model", the only difference in different approaches is on which level of abstraction the split is made between the two. On the highest level a shared model would be the requirements of the software system. Test and implementation is usually split at this level and that make it more likely to find faults in the implementation. 

Some interaction models that an exploratory test agent could use are:
\begin{itemize}
    \item An existing GUI (implicit model)
    \item A model artifact such as an FSM (explicit model)
    \item Other API describing artifacts such as Swagger documentation (explicit interface, implicit interactions)
\end{itemize}

There is a trade-off between implicit vs. explicit interaction models. With an explicit model the agent does not have to spend time in learning how to interact with the SUT, but developer time must be spent on building the model. On the other hand, with an implicit model the developer effort is lower but the time the agent must spend finding out how to interact is increased. This work could then replace manual work of explicit modeling with automatic implicit modeling.

\subsection{Agent Goals}
Depending on the interaction model, goals on different levels of exploration can be defined. The commonality of the goals is that they are on the level of abstraction of a human tester i.e. acceptance level and not on a unit- or integration-level.

It is also worth noting that learning more about the SUT can be a goal in itself. If a test suite passed yesterday and it also passes today the only thing we know is that we did not break anything that we tested for, given that the system is deterministic. No additional knowledge have been gained.

In cases with a model in the form of an FSM, one agent's goal could be to cover all the states in the FSM. Such agents would only fill the role of a model-based test, but on a system level, a group of such agents could apply learning of which different approaches to exploring the model result in new insights of the SUT.

In the case of using the GUI of an application, the interaction model is implicit in the controls of the GUI. For example, the agent would not be able to select an item in a list if such item has not first been created. With this kind of interaction model an agent's goals could be formulated as getting UI controls in a specific state or to get the complete UI looking in a certain way. In the first case a goal might be "get 10 items in this list" while in the latter, image analysis might be leveraged. An additional output and goal of such an exploration could be an artifact expressing the explicit model. This could enable the agent to visualize the interaction model and allow for a human expert to notice gaps or invalid states. 

Goals can also be on the system meta-level. Examples of meta-level agents are, an agent that observes the logs of a system, agents that stops system services and observes that the system gracefully recovers. Performance and resource consumption are other examples of such goals.

\subsection{Agent Assertions}
While agents run test cases their results are recorded. A part of the research challenges in this work is how such results should be recorded and how to generate oracles to assert on the result.

Ideally assertions should be easily formulated by developers and testers. Since the agents are exploring, assertions must take differences between tests in consideration. Given that, we think that good exploratory assertions will be property based. An example of such a property for an agent that have explored a GUI application, could be "The number of items in this list should be the difference between the times we have performed the 'create' state and the 'delete' state".
In practise, such a property would be validated by making a query of the test result and running that result through the list of assertions.

The agents should also have the capacity of asserting on unexpected system behaviour. Given results from previous tests an agent could learn and notify on any deviance.

\section{Empirical studies}

To evaluate the validity of this research the concepts and principles proposed will be deployed and empirically studied in an industrial setting on real products in development.

Currently we have several areas of activity on a real software system that is being developed. The system consists of multiple services that enable a mobile App client to view and subscribe to data from a process automation system.
The mobile App is adapted to enable a white-box interaction model in addition to a black-box UI based model. These can be evaluated as agent interaction models.

Upcoming studies will focus on the capability of the proposed methods to automatically detect faults and derive knowledge from this system under test.

\section{Current status of the research}

This research is in an early stage. For the back-end, a smoke-test service has been developed that run continuously in a staging environment. This service could be a candidate to make it into an agent with goals and more formalized collection of results.
Based on the systems Swagger documented API as interaction model, we are developing automatic property-based test case generation. We see this as a first step toward a more intelligent exploratory agent.

Where applicable any results in the form of software artifacts that are not proprietary could be shared as open-source software and presented at practitioner conferences.

\section{Conclusion}
In this paper we have proposed exploratory test agents as a way to increase the quality of software systems and to easier be able to reason about the SUT. In recent years several new tools have been proposed that uses some exploratory notions. However, these tools are usually implemented for a specific area and more work is needed to generalize between software systems.
The challenges identified to make exploratory test agents usable in industry and that need to be solved are, the interaction model with the SUT, the formulation of agent goals and how to assert on the results produced by the agents.
To find the answers to these questions we will conduct experiments in industry on real software systems.

\bibliographystyle{ACM-Reference-Format}
\bibliography{main}
\end{document}